\documentclass[sigconf,anonymous=false]{acmart}

\usepackage{booktabs} 
\usepackage{algorithm2e}

\AtBeginDocument{%
  \providecommand\BibTeX{{%
    \normalfont B\kern-0.5em{\scshape i\kern-0.25em b}\kern-0.8em\TeX}}}

\copyrightyear{2020}
\acmYear{2020}
\acmPrice{}

\acmConference[WWW'2020]{The Web Conference 2020}{April 20-24, 2020}{Taipei}
\acmYear{2020}
\copyrightyear{2020}

\begin{document}
\title{Solving Billion-Scale Knapsack Problems}

\author{Xingwen Zhang, Feng Qi, Zhigang Hua and Shuang Yang}
\affiliation{
  \institution{Ant Financial Services Group, San Mateo, CA 94402}
}
\email{{xingwen.zhang, feng.qi, z.hua, shuang.yang}@antfin.com}

\begin{abstract}
Knapsack problems (KPs) are common in industry, but solving KPs is known to be NP-hard and has been tractable only at a relatively small scale. This paper examines KPs in a slightly generalized form and shows that they can be solved \emph{nearly optimally} at scale via distributed algorithms. The proposed approach can be implemented fairly easily with off-the-shelf distributed computing frameworks (e.g. MPI, Hadoop, Spark). As an example, our implementation 
leads to one of the most efficient KP solvers known to date -- capable to solve KPs at an unprecedented scale (e.g., KPs with 1 billion decision variables and 1 billion constraints can  be solved within 1 hour). The system has been deployed to production and called on a daily basis, yielding significant business impacts at Ant Financial.
\end{abstract}

\ccsdesc[500]{Information systems~Information systems applications}

\keywords{Knapsack Problems; Large-scale Optimization; Distributed Algorithms; MapReduce; Dual Descent; Synchronous Coordinate Descent}

\maketitle

\section{Introduction}
\emph{Knapsack problems} (KPs) \cite{Kellerer2004} are commonly seen in real-world applications,
for example, budget allocation and pacing (as in advertising and marketing), online traffic control (as in search engine and recommender systems), logistics optimization (as in e-commerce), asset management (as in finance) and so on. Unfortunately, solving KPs is well known to be NP-hard and in practice only feasible at a relatively small scale even with commercial solvers \cite{Gurobi, CPlex, Agarwal2012, Gupta2016, Gupta2017, Zhao2018}.

We are primarily motivated by KPs as emerged in an internet industry setting, where decisions need to be made on a per user basis while the number of users can be large (e.g., billions). The ``resources" to be allocated in a KP can be either financial (e.g., loans, marketing promotions, ads spending, asset portfolios) or non-monetary (such as user impressions, clicks, dwell time). Typically, we want to optimize an \textit{objective} (e.g., the expected user conversions as in the case of marketing campaign) subject to a set of constraints, which can be roughly divided into two types: \textit{global constraints} that often limit the maximum allowance of the resources at a global level, and \textit{local constraints} that impose further restrictions specific to individual users / user groups.  
In practise, while the size of the global constraints are often small (e.g. a few hundreds), the typical scales for both the decision variables and the local constraints can be at the level of billions. Unfortunately, solving KPs at such scales has been an open technical challenge  \cite{Gurobi, CPlex, Agarwal2012, Gupta2016, Gupta2017, Zhao2018}.


We present in this paper one of the first attempts to solve real-world KPs at billion-scale. Firstly, using the \emph{MapReduce} computing model, we design a distributed framework for solving KPs by exploiting the decomposability of the dual problems and \emph{dual descent} (DD). Secondly, to further improve convergence especially as the DD algorithm is prone to constraint violations when implemented in a distributed setting, we developed the \emph{synchronous coordinate descent} (SCD) algorithm that doesn't suffer from these issues. Furthermore, by leveraging the hierarchical structures of the local constraints, we show that the integer programming sub-problem can be solved \emph{optimally in polynomial time by a greedy algorithm}, dramatically improving the quality and solving speed for KPs with hierarchical local constraints. Lastly, we implement our algorithm with off-the-shelf distributed computing frameworks (e.g. MPI, Hadoop, Spark), leading to one of the most efficient KP solvers known to date (e.g., KPs with 1 billion decision variables and 1 billion constraints can be solved within 1 hour). Our work also contributes to a deployed system that's being used for production decision making at Ant Financial everyday.

\section{Problem Formulation} \label{ProblemFormulation}
Consider the following \emph{generalized} variant of the knapsack problem (\ref{KPObjective})--(\ref{KPFormulationEnd}),
where a set of $M$ items are to be allocated among a set of $N$ users, subject to $K$ global constraints and $L$ local constraints. The global constraints \eqref{GlobalConstraints} limit the resource allocation for each \emph{knapsack}, whereas the local constraints \eqref{LocalConstraints} restrict per-user consumption. If item $j$ is allocated to group\footnote{Throughout this paper, the terms ``user'' and ``group'' are used interchangeably, as a ``user'' in our setting corresponds to a ``group'' in the \emph{operations research} literature.} $i$, i.e. $x_{i,j} = 1$, we gain a profit of $p_{i,j}$ and consume  $b_{i,j,k}$ amount of resources for the $k$-th knapsack ($k\in[K]=\{1,\dots,K\}$). $B_k$ and $C_l$ are \emph{strictly positive} while $p_{i,j}$ and $b_{i,j,k}$ are \emph{non-negative}.

\begin{flalign}
\max_{x_{i,j}}&\displaystyle \sum_{i=1}^{N}\sum_{j=1}^{M}{p_{i,j}x_{i,j}} \label{KPObjective} \\
s.t.& \displaystyle \sum_{i=1}^{N}\sum_{j=1}^{M}{b_{i,j,k}x_{i,j}} \leq B_k, \ \forall k \in [K] \label{GlobalConstraints} \\
& \displaystyle \sum_{j\in S_l}x_{i,j} \leq C_l, \ \forall i \in [N], S_l \subseteq [M], \forall l \in [L]  \label{LocalConstraints} \\
& x_{i,j} \in \{0,1\}, \forall i \in [N], \forall j \in [M]. \label{KPFormulationEnd}
\end{flalign}
Note that although we assume  $x_{ij}$'s are \emph{binary} (i.e., $x_{ij}\in \{0, 1\}$), our approach is equally applicable to \emph{categorical} (i.e., non-negative integer) variables. In fact, our implementation supports both binary and categorical decision variables.



\subsection{Hierarchical Local Constraints}


We further explore a more complex case of (\ref{KPObjective})--(\ref{KPFormulationEnd}), where there are hierarchical structures in the local constraints (\ref{LocalConstraints}) such that the index sets are \emph{either} disjoint \emph{or otherwise} nested. Formally,



\begin{definition}
Local constraints (\ref{LocalConstraints}) are said to be \emph{hierarchical} when the following holds: $\forall l,l'\in[L]$, if $S_l \cap S_{l'} \neq \emptyset$, then \emph{either}  $S_l \subseteq S_{l'}$ \emph{ or } $S_{l'}\subseteq S_{l}$.
\end{definition}

This property is commonly seen in real-world, where items are not independent from each other but rather related and often the time organized as nested groups (e.g., according to a taxonomy). A \emph{directed acyclic graph} (DAG) can be constructed for $\{S_l|l\in[L]\}$, i.e., an arc from $S_l$ to $S_{l'}$ iff $S_l \subseteq S_{l'}$. In the trivial case where items are unrelated, this DAG will degenerate to a \emph{set} where $S_l$'s are disjoint from each other.

\subsection{Connections to Other KP Variants}

A few variants of the knapsack problems have been studied in the literature, including \emph{multi-dimensional knapsack problems} (MDKPs) \cite{Chu1998}, \emph{multi-choice knapsack problems} (MCKPs) \cite{Kellerer2004}, and \emph{multi-dimensional multi-choice knapsack problems} (MMKPs) \cite{Khan1998}. An MDKP has a single item and multiple knapsack constraints, when the item is chosen, resources from multiple knapsacks will be consumed. MCKP is an extension of the classical single-constraint KP, where the items are partitioned into multiple groups and exactly one item from each group can be chosen. MMKP is a combination of MDKP and MCKP \cite{Khan1998, Htiouech2013}.

The problems we study here (\ref{KPObjective})--(\ref{KPFormulationEnd}), including the case with the local hierarchical assumption, are more generalized compared to the other variants as they allow more flexible forms of constraints. In fact, \textit{all} these classical variants can be seen as special cases of our formulation. For example, if there's only one item and no local constraint (i.e., $M=1$ and $L=0$), it reduces to MDKPs \cite{Chu1998}; when $K=1, C_{\cdot} =1$ and $L=1$, it becomes MCKPs \cite{Kellerer2004}; and when $C_{\cdot} =1$ and $L=1$, MMKPs \cite{Khan1998}.

\section{Related Work}\label{Related}

As a famous example of the \emph{integer programming} (IP) problem, KPs (including vanilla KP and its variants) are well-known to be NP-hard \cite{Martello1990, Kellerer2004, Khan1998}. Both exact and heuristic algorithms have been studied for solving these problems in operations research (OR), including, for example, \emph{branch and bound} \cite{Chu1998}, \emph{tabu search} \cite{Glover1989}, \emph{simulated annealing search} \cite{Kirkpatrick1983} and so on. See \cite{Laabadi2018} for a recent survey on solving MDKPs. Unfortunately, these traditional OR algorithms were not designed for modern infrastructure, in particular distributed computing. As a result, they can only solve KPs at a very limited scale (i.e., thousands to millions of decision variables).

The recent works by Pinterest \cite{Zhao2018} and LinkedIn \cite{Gupta2016, Gupta2017} are the only few cases that have examined KPs at a scale close to ours. In \cite{Zhao2018}, a simplified KP is solved to decide notification volume for each user so as to optimize long-term user engagements. Their \emph{threshold search} algorithm is able to solve KPs at the scale of hundreds of millions, but only when there is a single global constraint. In \cite{Gupta2016, Gupta2017}, the authors formulated the email volume optimization problem at LinkedIn as a multi-objective \emph{linear programming} (LP) problem, which is converted to a strongly convex dual problem with an added quadratic regularization term as in \cite{Agarwal2012}, and the set of dual multipliers is used for online production serving. To make the solution tractable, clustering or sampling techniques are used to reduce the number of decision variables \cite{Agarwal2012}. The LP relaxation as well as the downsizing procedure, however, inevitably hurt the quality of the resultant solution (i.e., optimality and constraint satisfaction).

\section{Our Approach} \label{MapReduceSection}
The scale of the problems we aim to solve suggests that the solution ought to be computed by a distributed cluster rather than a single machine. In this section, we first introduce the dual decomposition of KPs that opens the door to distributed solving, and then develop distributed solving algorithms described using the \emph{MapReduce} computing model\footnote{The MapReduce semantics are used here only to describe the algorithm. The implementation, however, can be based on any other distributed computing frameworks (e.g., MPI, Hadoop, Spark).} \cite{Dean2008}.

\subsection{\textbf{Dual Decomposition for KPs}}
Let us examine the dual problem of (\ref{KPObjective})--(\ref{KPFormulationEnd}). By introducing a set of Lagrangian multipliers\footnote{In economics, $\lambda_k$ is often interpreted as the \emph{shadow price} (or \emph{marginal utility}) of the $k$-th knapsack resource \cite{Boyd2004}.} $\{\lambda_k|k \in[K] \}$ (i.e., one for each global constraint), we have
\begin{flalign}
\max_{x_{i,j}}&\sum_{i=1}^{N}\sum_{j=1}^{M}{p_{i,j}x_{i,j}} - \sum_{k=1}^{K}{ \lambda_k (\sum_{i=1}^{N}\sum_{j=1}^{M}{b_{i,j,k}x_{i,j}} - B_k)} \label{MaxFunction}
\\
s.t.& \sum_{j\in S_l}{x_{i,j}} \leq C_l, \ \forall i\in [N], S_l\subseteq [M], \forall l \in [L]  \label{LocalConstraintsInPrimal}
\\
& x_{i,j} \in \{0,1\}, \forall i \in [N],  \forall j \in [M] \label{MaxFunctionEnd}
\end{flalign}
with optimality conditions
\begin{flalign}
&\lambda_k (\displaystyle \sum_{i=1}^{N}\sum_{j=1}^{M}{b_{i,j,k}x_{i,j}} - B_k) = 0, \forall k \in [K] \label{EquationKKTStart} \\
&  \sum_{i=1}^{N}\sum_{j=1}^{M}{b_{i,j,k}x_{i,j}} - B_k \leq 0, \forall k \in [K]  \label{EquationPrimalFeasibility} \\
& \lambda_k \geq 0, \forall k \in [K]. \label{EquationKKTEnd}
\end{flalign}

The maximization problem in (\ref{MaxFunction}) can be decomposed into a set of subproblems (i.e., one for each group $i$),
\begin{flalign}
\max_{x_{i,j}}& \sum_{j=1}^{M}{p_{i,j}x_{i,j}} - \sum_{k=1}^{K}{ \lambda_k \sum_{j=1}^{M}{b_{i,j,k}x_{i,j}} }  \label{MaxFunctionPerGroup} \\
s.t.&  \sum_{j\in S_l}x_{i,j} \leq C_l, S_l\subseteq [M], \ \forall l \in [L] \label{LocalConstraintsPerGroup} \\
& x_{i,j} \in \{0,1\}, \forall j \in [M]. \label{BinaryConstraintsPerGroup}
\end{flalign}

Given $\lambda$s, these subproblems are independent from each other. This nice decomposability of the dual suggests a natural decentralized approach for solving large-scale KPs, i.e., by alternating between (1) solving the IP subproblems \emph{in parallel} with given $\lambda$; and (2) updating $\lambda$ while fixing the solution $x$.

\subsection{Solving the IP Subproblems \label{gis}}
Thanks to the decomposability of the dual, at a fixed $\lambda$, KPs can be solved at the individual user level \emph{independently in parallel}. Compared to the original KPs, the per-user IP subproblems (\ref{MaxFunctionPerGroup})-(\ref{BinaryConstraintsPerGroup}) are much easier to solve as (1) the scale is tiny -- there're only $O(M)$ decision variables (\emph{vs} $O(MN)$ of the original KPs); and (2) the problem is much simpler (e.g., there's no global constraint). In practice, it's straightforward to solve these subproblems, e.g., by bundling an off-the-shelf solving subroutine (e.g., \cite{ORTools, Gurobi, CPlex}) into the deployable image of the mapper.

For the more complex case with hierarchical local constraints, there's a fast algorithm that has a polynomial time complexity and is provably optimal. In particular, we design a greedy algorithm, as described in Algorithm \ref{AlgorithmGreedy}, which solves the IP subproblems by traversing the DAG in a topological order. The algorithm orders items in a non-decreasing order of \textit{cost-adjusted profit} (which is also the contributing dual value of $x_{i,j}$),
\begin{eqnarray*}
\tilde{p}_{i,j} = p_{i,j} - \sum_{k=1}^{K}{\lambda_k b_{i,j,k}}.
\end{eqnarray*}
 Starting from the items at the lowest level of the DAG, for each $S_l$ the algorithm chooses its items in a non-decreasing order of the cost-adjusted profit until their sum exceeds $C_l$. The $x_{i,j}$ for the unchosen items in $S_l$ are all assigned with value 0 and these items will not be considered in the subsequent iterations. This greedy selection process is repeated until all the nodes in the DAG have been traversed.

\begin{algorithm}
\SetKwFor{ForEach}{for each}{do}{end}
\SetArgSty{textrm} 
\DontPrintSemicolon
\SetAlgoLined
Initialize for $j\in[M]$: $x_{i,j} = 1 \text{ if } \tilde{p}_{i,j} > 0, \text{ or } 0 \text{ otherwise}$ \;
Sort $\{j\}$ in non-increasing order of $\tilde{p}_{i,j}$ \;
\For{$S_l$ in the topological order of the DAG} {
    fetch the indices $j \in S_l$ with $x_{i,j}=1$ \;
    update $x_{i,j}=0$ if $j$ is not in top $C_l$ sorted indices
}
\Return $\{x_{i,j}|j\in[M]\}$ \;
\caption{Greedy algorithm for solving the per-user IP subproblem (\ref{MaxFunctionPerGroup})--(\ref{BinaryConstraintsPerGroup}) with local hierarchical constraints.}
\label{AlgorithmGreedy}
\end{algorithm}

This greedy algorithm has a polynomial time complexity, and as is shown later it's orders of magnitude faster than competitive solvers. We also prove it is optimal.

\begin{proposition}
Algorithm \ref{AlgorithmGreedy} optimally solves the integer programming problem (\ref{MaxFunctionPerGroup})--(\ref{BinaryConstraintsPerGroup}) with hierarchical local constraints.
\end{proposition}

\begin{proof} Given any other solution (denoted by $\{\tilde{x}_{i,j}, j\in[M]\}$) that satisfies the constraints (\ref{LocalConstraintsPerGroup}) and (\ref{BinaryConstraintsPerGroup}) but differs from the greedy solution (denoted by $\{x^{*}_{i,j}, j\in[M]\}$), we can identify the first node in the topological order of the DAG at which the items chosen are different. Due to the nature of the greedy algorithm, there must exist a pair of items $j$ and $j'$ at the node where the adjusted profit of item $j$ is no less than that of item $j'$, but $x^{*}_{i,j}=1, x^{*}_{i,j'}=0$, $\tilde{x}_{i,j}=0, \tilde{x}_{i,j'}=1$. We can modify $\tilde{x}$ by setting $\tilde{x}_{i,j}=1, \tilde{x}_{i,j'}=0$ without decreasing the objective value of (\ref{MaxFunctionPerGroup}). All the constraints (\ref{LocalConstraintsPerGroup}) and (\ref{BinaryConstraintsPerGroup}) are still satisfied, because any later node in the topological order of the DAG contains both $j$ and $j'$ or neither. In this way, we can always convert any solution to the greedy solution without decreasing the objective value or violating any constraint. This completes the proof.
\end{proof}

\subsection{Distributed Algorithms for Solving KPs}

\subsubsection{Dual Descent}
Our first algorithm for solving KPs is distributed \emph{dual descent} (DD). It's an iterative procedure developed based on the decomposability of the dual. Given the solutions of the per-user subproblems, $\{x_{ij}\}$, the multipliers $\lambda$ can be updated:
\begin{eqnarray}
\lambda_k^{t+1} = \max(\lambda_k^{t} + \alpha ( \sum_{i=1}^N\sum_{j=1}^M b_{i,j,k}x_{i,j} - B_k), 0),
\label{EquationGradientDescent}
\end{eqnarray}
where the hyper-parameter $\alpha$ is the learning rate.
In Algorithm \ref{AlgorithmIterative}, we detail it using the MapReduce model \cite{Dean2008}. Particularly, in each iteration: 1) first, the solutions $x_{i,j}$ for the subproblems are computed independently in mappers (e.g., by off-the-shelf IP solvers \cite{Gurobi,CPlex,ORTools}, or in the hierarchical case by our greedy algorithm); once finished, each mapper emits $K$ values $\{v_{ik} =\sum_{j=1}^{M}b_{i,j,k}x_{i,j}|\ k\in[K]$\} corresponding to group $i$'s resource consumption from each of the $K$ knapsacks; 2) then, the reducers aggregate the total resource consumption of each knapsack, $R_k = \sum_{i=1}^{N}v_{ik}$; and 3) finally, a master node updates $\lambda_k$ using dual descent (\ref{EquationGradientDescent}).


\begin{algorithm}
\SetKwFor{ForEach}{for each}{do}{end}
\SetArgSty{textrm} 
\SetKwFunction{Map}{Map}
\SetKwFunction{Reduce}{Reduce}
\SetKwProg{myproc}{Procedure}{}{}
\DontPrintSemicolon 
\myproc{\Map{$p_i$, $b_i$, $\lambda^t$}}{
    solve $x_{i,j}$ from (\ref{MaxFunctionPerGroup}) \;
    \ForEach{$k \in \{1,\dots,K\}$} {
        emit($k$, $v_{ik} = \sum_{j=1}^{M}{b_{i,j,k}x_{i,j}}$)\;
    }
}
\;
\myproc{\Reduce{$k$, $v_{ik}$}} {
    \Return $\sum_i{v}_{ik}$ \;
}
\SetAlgoLined
\;
initialize $\lambda^0$ \;
\For{$t \gets 0$ \KwTo $T - 1$} {
    \ForEach{$i \in \{1,\dots,N\}$ in parallel} {
        \Map{$p_i$, $b_i$, $\lambda^t$} \;
    }
    \ForEach{$k \in \{1,\dots,K\}$ in parallel} {
        $R_k$ = \Reduce{$k$, $v_{ik}$} \;
    }
    \ForEach{$k \in \{1,\dots,K\}$} {
        $\lambda_k^{t+1} = max(0$, $\lambda_k^{t} + \alpha (R_k - B_k)$) \;
    }
    \lIf{$\lambda^{t}$ has converged}{\Return $\lambda^t$}
}
\Return $\lambda^T$
\caption{Distributed dual descent algorithm for solving KPs.}
\label{AlgorithmIterative}
\end{algorithm}

\subsubsection{Synchronous Coordinate Descent} \label{scd}
 This DD algorithm, however, is problematic because: (1) it requires a hyper-parameter $\alpha$ that needs to be chosen either manually or programmatically, which can be practically cumbersome or computationally intensive especially for large-scale KPs; and (2) as we show empirically, DD's very prone to constraint violations and as a result the resultant solutions are often invalid. To this end, we propose to use \emph{coordinate descent} (CD), which updated $\lambda_k$ one coordinate at a time while keeping other coordinates $\{\lambda_{k'}|k'\neq k\}$ fixed.

The CD algorithm can be further enhanced for the hierarchical case. Specifically, we note that it's possible to dramatically reduce the search space for $\lambda$ by only considering a small set of candidate values. To show this, observe from Algorithm \ref{AlgorithmGreedy} that the solution only depends on the relative order of the cost-adjusted profits, $\tilde{p}_{i,j}$, rather than their actual values. For any given $i$, let $z_{j,k} = (p_{i,j} - \sum_{k'=1, k' \neq k}^{K}{\lambda_{k'} b_{i,j,{k'}}}) - \lambda_{k} b_{i,j,{k}}$ denote a linear function $z_{j,k}(\lambda_k)$, which defines a straight linear in the $(\lambda_k, z_{j,k})$-axis. The relative order of $\tilde{p}_{i,j}$ can only possibly change \emph{either} at (1) the pairwise intersections of these $M$ lines (i.e., $\{\lambda_k | z_{j,k}(\lambda_k) = z_{j'\neq j,k}(\lambda_k), \forall j,j'\in[M]\}$), \text{or} at (2) their intersections with the horizontal axis (i.e., $\{\lambda_k | z_{j,k}(\lambda_k) = 0, \forall j\in[M]\}$). Therefore, it suffices to only screen $\lambda_k$ at these values instead of over the entire interval $[0,\infty)$. Algorithm \ref{AlgorithmInteract} describes this procedure for deciding potential candidate values for $\lambda_k$.

\begin{algorithm}
\SetKwFor{ForEach}{for each}{do}{end}
\SetArgSty{textrm} 
\SetKwFunction{Intersect}{Intersect}
\SetKwFunction{CalculateIntersectionPoints}{CalculateIntersectionPoints}
\SetKwFunction{Map}{Map}
\SetKwFunction{Reduce}{Reduce}
\SetKwFunction{Optimize}{Optimize}
\SetKwProg{myproc}{Procedure}{}{}
\DontPrintSemicolon
\myproc{\CalculateIntersectionPoints{$p_i$, $b_i$, $\lambda^t$, $k$}} {
    $\Lambda_k$ = $\emptyset$ \;
    \ForEach{$j \in  \{1,\dots,M\}$} {
        $\Lambda_k$ = $\Lambda_k\cup\{\lambda_k | z_{j,k}(\lambda_k) = 0\}$\;
        \ForEach{$j' \in \{j+1,\dots,M\}$} {
            $\Lambda_k$ = $\Lambda_k\cup \{\lambda_k | z_{j,k}(\lambda_k) = z_{j',k}(\lambda_k)\}$  \;
        }
    }
    \Return the set of unique values in $\Lambda_k$
}
\caption{Calculating candidate values for $\lambda_k$ (in parallel for each group $i$) for KPs with hierarchical local constraints.}
\label{AlgorithmInteract}
\end{algorithm}

Once we have the candidate $\lambda$ values, for each group $i$ \emph{independently in parallel}, we can update $\lambda$ using \emph{synchronous coordinate descent} (SCD), as described in Algorithm~\ref{AlgorithmSCD}.  Specifically, for each group $i$, the mapper goes through the candidate values for $\lambda_k$ and calculates the amount of $k$-th knapsack resource that would be used if we update $\lambda_k$ to the corresponding candidate value. The mapper sorts the candidates of new $\lambda_k$ in a decreasing order and emits only the \textit{incremental} change as we decrease $\lambda_k$. For each global constraint $k$, the reducer then aggregates the emitted results for each key, and then $\lambda_k$ is updated to the minimal threshold such that the total resource used does not exceed $B_k$.

While we employ synchronous CD that updates $\lambda_k$ for all $k\in[K]$ simultaneously, other variants of CD, such as \emph{cyclic} CD (updates one multiplier at a time) and \emph{block} CD (updates multiple multipliers in parallel) are also applicable \cite{Wright2015}. In our implementation, all these modes are supported, although SCD turns to perform the best.

\begin{algorithm}
\SetKwFor{ForEach}{for each}{do}{end}
\SetArgSty{textrm} 
\SetKwFunction{Intersect}{Intersect}
\SetKwFunction{CalculateIntersectionPoints}{CalculateIntersectionPoints}
\SetKwFunction{Map}{Map}
\SetKwFunction{Reduce}{Reduce}
\SetKwFunction{Optimize}{Optimize}
\SetKwProg{myproc}{Procedure}{}{}
\DontPrintSemicolon
\myproc{\Map{$p_i$, $b_i$, $\lambda^t$}} {
    \ForEach{$k \in \{1,\dots,K\}$} {
        $\Lambda_k$ = \CalculateIntersectionPoints{$p_i$, $b_i$, $\lambda^t$, $k$} \;
        Sort $\Lambda_k$ in a decreasing order \;
        previous\_sum = 0 \;
        \ForEach{$\lambda_k$ in $\Lambda_k$} {
            $\lambda=(\lambda_1^t,\dots,\lambda_{k-1}^t,\lambda_k,\lambda_{k+1}^t,\dots,\lambda_{K}^t)$ \;
            solve $x_{i,j}$ with multipliers $\lambda$ using Algorithm \ref{AlgorithmGreedy} \;
            current\_sum = $ \sum_{j=1}^{M}{b_{i,j,k}x_{i,j}}$ \;
            \If{current\_sum > previous\_sum}{
                $v_1$ = $\lambda_k$ \;
                $v_2$ = current\_sum - previous\_sum \;
                emit($k$, [$v_1$, $v_2$]) \;
            previous\_sum = current\_sum \;
            }
        }
    }
}
\;
\myproc{\Reduce{k, [$v_1$, $v_2$]}} {
    \eIf{$\sum{v_2} \leq B_k$}{
        \Return 0
    }{
        \Return minimal threshold $v$ such that $\displaystyle \sum_{v_1 \geq v}{v_2} \leq B_k$
    }
}
\;
\SetAlgoLined
initialize $\lambda^0$ \;
\For{$t \gets 0$ \KwTo $T - 1$} {
    \ForEach{$i \in \{1,\dots,N\}$ in parallel} {
        \Map{$p_i$, $b_i$, $\lambda^t$} \;
    }
    \ForEach{$k \in \{1,\dots,K\}$ in parallel} {
         $\lambda_k^{t+1}$ = \Reduce{$k$, [$v_1$, $v_2$]} \;
    }
    \Return $\lambda^t$ if {$\lambda^{t}$ is converged}
}
\caption{Synchronous coordinate descent for solving KPs.}
\label{AlgorithmSCD}
\end{algorithm}

\subsection{\textbf{Convergence}}

For the special case of $K=1$, it is straightforward to show that our Algorithm \ref{AlgorithmSCD} converges to a solution whose total profit is at most $\max_{i,j}{p_{i,j}}$ less than the optimal solution (because it produces a rounded integer solution of a \emph{fractional knapsack problem}) \cite{Dantzig1957}.

For more general cases, if the algorithm converges to a pair ($x^{*},\lambda^{*}$) that jointly satisfy (\ref{MaxFunction})--(\ref{EquationKKTEnd}) (i.e., so-called ``sufficient'' optimality conditions), $x^{*}$ is then optimal for the primal problem \cite{Shapiro1979}. However, there is no theoretical guarantee on the convergence of the algorithm to such a pair ($x^{*},\lambda^{*}$). In fact, the solution $x$ computed for the maximization problem (\ref{MaxFunction})--(\ref{MaxFunctionEnd}) is not guaranteed to be even feasible for the problem (\ref{KPObjective})--(\ref{KPFormulationEnd}). Nevertheless, as shown in \cite{Shapiro1979}, the solution $x$ computed for (\ref{MaxFunction})--(\ref{MaxFunctionEnd}) is optimal for a family of IP problems if $B_k$ is replaced by  $\sum_{i=1}^{N}\sum_{j=1}^{M}{b_{i,j,k}x_{i,j}} + \delta_k$ where $\delta_k$ is non-negative (equals to 0 when $\lambda_k > 0$). We analyze the optimality gap empirically and show our algorithms are often nearly optimal. In particular, the optimality gap decreases as the number of users $N$ increases. When $K \ll N$ (which is often the case in real-world applications \cite{Zhao2018,Gupta2016, Gupta2017}), the optimality gap is so small that the resultant solution is very close to optimal.

\section{Further Speedups} \label{ExpeditionSection}

\subsection{\textbf{Linear-time $\lambda$ Candidate Generation}}
The calculation of candidate $\lambda$ values for each group in the generalized algorithm has a time complexity of $O(KM^{3}\log{M})$. This can be further enhanced when global constraints are in \textit{sparse} form and one local constraint exists per group such that:
\begin{itemize}
    \item there exists a one-to-one mapping between the items and the knapsacks (i.e., $M=K$, and $b_{i,j,k} = 0, \forall j \neq k$), and
    \item there is one local constraint limiting the maximum number (denoted by $Q$ thereafter) of items chosen for each group.
\end{itemize}
For such cases, there is \emph{at most} one candidate for $\lambda_k$ depending on whether the $k$-th item has a top $Q$ adjusted profit or not. In particular, if the adjusted profit of item $k$ is already in top-$Q$ list, the critical value of new $\lambda_k$ is the one that lowers its adjusted profit to the $(Q+1)$-th adjusted profit. If the adjusted profit of item $k$ is not in top $Q$, the critical value of new $\lambda_k$ is the one that increases its adjusted profit to the $Q$-th adjusted profit. The pseudo-code is given in Algorithm \ref{AlgorithmOKMap} where $\bar{p}$ is the threshold deciding whether the $k$-th item will be chosen for group $i$. If new $\lambda_k$ is larger than $\frac{p_{i,k} - \bar{p}}{b_{i,k,k}}$, the updated adjusted profit of the $k$-th item will be below $\bar{p}$ and thus the item will not be chosen. On the other hand, a new $\lambda_k$ below $\frac{p_{i,k} - \bar{p}}{b_{i,k,k}}$ guarantees that the resulting adjusted profit of the $k$-th item is among top $Q$ across all items. Thus, Algorithm \ref{AlgorithmOKMap} correctly emits the only candidate of new $\lambda_k$ (if any) that determines whether the $k$-th item has a top $Q$ adjusted profit or not.
\begin{algorithm}
\SetKwFunction{Map}{Map}
\SetKwProg{myproc}{Procedure}{}{}
\DontPrintSemicolon
\myproc{\Map{$p_i$, $b_i$, $\lambda^t$}} {
    initialize adjusted\_profits as an array of $K$ numbers \;
    \ForEach{$k \in \{1,\dots,K\}$} {
        adjusted\_profits[$k$] = max($p_{i,k} - \lambda_k^t b_{i,k,k}, 0$)
    }
    $Q$\_th\_largest = quick\_select(adjusted\_profits, $Q$) \;
    $Q1$\_th\_largest = quick\_select(adjusted\_profits, $Q + 1$)\;
    \ForEach{$k \in \{1,\dots,K\}$} {
        \eIf{adjusted\_profits[$k$] $\geq$ $Q$\_th\_largest}{
            $\bar{p}$ = $Q1$\_th\_largest \;
        } {
            $\bar{p}$ = $Q$\_th\_largest \;
        }
        \If{$p_{i,k} > \bar{p} $}{
            $v_1 =  \frac{p_{i,k} - \bar{p}}{b_{i,k,k}}$\;
            $v_2 = b_{i,k,k}$ \;
            emit($k$, [$v_1$, $v_2$]) \;
        }
    }
}
\caption{Linear time $Map$ function for choosing up to $Q$ items, each corresponding to a knapsack.}
\label{AlgorithmOKMap}
\end{algorithm}

Algorithm \ref{AlgorithmOKMap} uses $quick\_select(array, n)$ to find the $n$-th largest element of an $K$-array. The overall complexity of this procedure is $O(K)$, independent of the value of $Q$ \cite{Cormen2009}.

\subsection{Fine-tuned Bucketing}
A straightforward implementation of the $Reduce$ function in Algorithm \ref{AlgorithmSCD} is to sort the emitted results by the value of $v_1$ and choose the minimal threshold $v$ based on the sorted results. A speedup is to bucket the values of $v_1$ and calculate the sum of $v_2$ for each bucket. We then identify the target bucket that the threshold $v$ falls into, and approximate the value of $v$, for example, by interpolating within the bucket.

To improve the accuracy of the above approximation through bucketing and interpolating, we would like that the bucket size is small around the true value of $v$ and large when the bucket is unlikely to contain $v$. Unfortunately, the true value of $v$ is unknown to us beforehand. Nevertheless, due to the iterative nature of Algorithm \ref{AlgorithmSCD}, the value calculated in the previous iteration, i.e. $\lambda_k^t$, provides a reasonable estimate for $v = \lambda_k^{t+1}$. We thus designed an uneven bucketing scheme such that the bucket is of a minimal size around $\lambda_k^t$ and grows exponentially as it deviates from $\lambda_k^t$. Specifically, given the value calculated in the previous iteration $\lambda_k^t$, the bucket id assigned to a candidate $\lambda_k$ is given as
\begin{eqnarray*}
bucket\_id(\lambda_k) = sign(\lambda_k - \lambda_k^t) \lfloor\log{\frac{|\lambda_k - \lambda_k^t|}{\Delta}}\rfloor,
\end{eqnarray*}
where $\Delta$ is a parameter controlling bucket sizes.

\subsection{Pre-solving by Sampling}
Like all other iterative algorithms, starting from a good initialization can significantly accelerate the convergence of the algorithm. The initial value for the dual multipliers, $\lambda^0$ in Algorithm \ref{AlgorithmSCD}, is often chosen randomly, but can be estimated by pre-solving using sampled data.  By sampling small sets of random groups and adjusting knapsack budgets proportionally, we can start the algorithm with better initialization. Our experiments show that pre-solving can save up to 40\% to 75\% of iterations for large-scale KPs.

\subsection{Post-processing for Feasibility} \label{SectionPostprocessing}
For a converged solution from our algorithms, it's likely that the total resource consumption violates the global constraints (\ref{GlobalConstraints}) just by a tiny bit. To ensure the satisfaction of the global constraints and also accelerate the convergence, we propose a light-weight post-processing method based on the \textit{cost-adjusted group profit} quantity,
\begin{eqnarray*}
\tilde{p}_i = \displaystyle \sum_{j=1}^{M}{ p_{i,j}x_{i,j}} - \sum_{k=1}^{K}{ \lambda_k \sum_{j=1}^{M}{b_{i,j,k}x_{i,j}} }.
\end{eqnarray*}
which is actually also the dual value contributed by a given group $i$. The post-processing procedure works by sorting the groups $\{i\}$ according to the non-decreasing order of $\tilde{p}_i$, and in this order one by one, resetting $x_{i}$ to 0 until all global constraints are satisfied. Since the cost-adjusted group profit measures the benefit of choosing items from group $i$, removing those with smaller values of $\tilde{p}_i$ until constraint satisfaction is a sensible heuristic to \emph{project} the solution to its nearest neighbor (a boundary point) in the feasible region.

\section{Experiments} \label{ExperimentsSection}
We tested the distributed solvers on both synthetic and real-world data\footnote{In our typical real-world applications there are hundreds of millions users. As long as we are aware, no datasets of comparable size are publicly available.}. Unless otherwise stated, for synthetic instances $p_{i,j}$ is uniformly distributed in $[0,1]$. Two classes of global constraints (\textit{sparse} and \textit{dense}) are experimented with, where the cost coefficient $b_{i,j,k}$ is uniformly sampled within $[0,1]$. Without losing generality, the budgets of global constraints are scaled with $M$, $N$ and $L$ to ensure tightness of constraints, and $C_l$) are set to 1.

In the following, \textit{optimality ratio} is defined as the ratio of primal objective value to relaxed LP objective value; \textit{constraint violation ratio} is defined as the ratio of excessive budget to given budget for a constraint, and we use \textit{max constraint violation ratio} over all constraints to quantify overall violation of a solution.

\subsection{Optimality Evaluation} \label{SectionOptimality}
We evaluate the optimality ratio between our KP solution against relaxed LP solution (i.e., the upper bound). It is difficult to find any existing LP solver that can handle billion-scale problems, therefore our comparison has to be restricted to modest-size problems so we can get the upper-bound using LP solvers such as Google OR-tools \cite{ORTools}.

\begin{figure}
\centering
\includegraphics[width=0.8\linewidth]{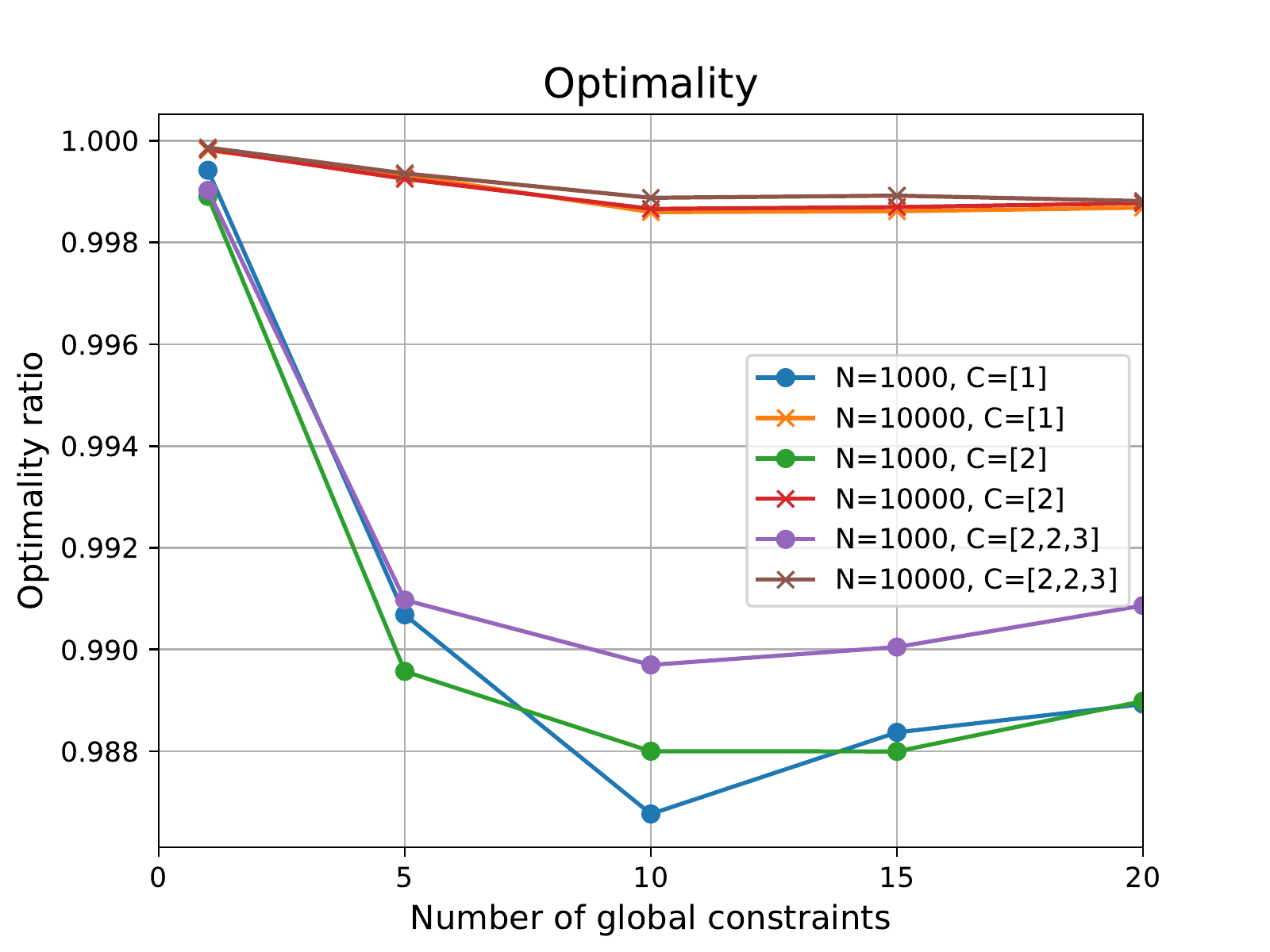}
\caption{Optimality ratio between KP solutions and upper bounds computed by LP relaxation}
\label{OptimalityGap}
\end{figure}

Figure \ref{OptimalityGap} shows the optimality ratios for $N=1,000$ and $10,000$ across $K \in \{1, 5, 10, 15, 20\}$. We fix the number of items per group $M$ at 10. To increase the diversity of the items, $b_{i,j,k}$ are uniformly distributed in $[0,1]$ and in $[0,10]$ with equal probabilities (each being 0.5). We evaluated optimality ratios for three scenarios of local constraints as shown in Figure \ref{OptimalityGap} where
\begin{itemize}
\item C=[1] means $\sum_{j=1}^{M}{x_{i,j}} \leq 1$,
\item C=[2] means $\sum_{j=1}^{M}{x_{i,j}} \leq 2$, and
\item C=[2,2,3] corresponds to hierarchical local constraints.
\end{itemize}
We plot the average optimality ratio (across 3 runs) as we vary $K$, as shown in Figure \ref{OptimalityGap}. The optimality gap decreases as $N$ increases. The optimality ratio is above 98.6\% for all experiment cases, and above 99.8\% for $N=10,000$ under all scenarios of local constraints.

\subsection{Duality Gap on Large Datasets}
We tested our system with large synthetic data to measure solution quality as well. The large-scale test sets contain a number of sparse global constraints with $N = 100$ million users, while the number of items $M$ in each group varies from 1 to 100 and thus the total number of items we consider is up to 10 billion. Table \ref{TableLargeData} shows the number of SCD iterations, primal objective values and duality gaps\footnote{A duality gap is the difference between the dual objective value and the primal IP objective value.}\cite{Boyd2004}. The duality gaps are much smaller than the primal objective values, indicating that the solutions are nearly optimal. Furthermore, no global constraint is violated when our algorithm converges.
\begin{table}
\caption{Results for 100 million users with up to 10 billion items}
\label{TableLargeData}
\centering
\begin{tabular}{lccccc}
\toprule
$M$ & Iterations & Primal value & Duality gap \\
\midrule
1 & 2 & 40,631,183.07 & 0.0 \\
5 & 13 & 73,000,742.85 & 302.71 \\
10 & 18 & 85,378,580.47 & 290.74 \\
20 & 14 & 92,415,786.11 & 301.04 \\
100 & 10 & 98,436,146.56 & 225.15 \\
\bottomrule
\end{tabular}
\end{table}

\subsection{Effectiveness of Pre-solving} \label{SectionPresolvingResults}
When the number of groups $N$ is large, pre-solving with sampled users is used to generate good starting points for $\lambda$. We sample $n=10,000$ groups for pre-solving, and the computation time of pre-solving is negligible since $n \ll N$. We compare the number of SCD iterations until convergence with pre-solving and without pre-solving, both starting at $\lambda_k=1.0, \forall k\in[K]$. Table \ref{TablePresolving} reports the number of SCD iterations for sparse problem instances with $N=1$ million, 10 million, 100 million. For each $N$, we fix $M=10$ and $K=10$. The results in Table \ref{TablePresolving} show that pre-solving reduced the number of SCD iterations by 40\% to 75\%.
\begin{table}
\caption{Number of SCD iterations with/without pre-solving}
\label{TablePresolving}
\centering
\begin{tabular}{cccc}
\toprule
$N$ & No pre-solving & Pre-solving & \% of reduction \\
\midrule
1 million & 35 & 21 & 40\% \\
10 million & 32 & 8 & 75\% \\
100 million & 32 & 13 & 59\% \\
\bottomrule
\end{tabular}
\end{table}

We also notice pre-solving alone is not sufficient for solving KP problems, as $\lambda$ produced by pre-solving led to constraint violations. When applying pre-solved $\lambda$ to full datasets, we observed that the number of global constraint violations are 4, 5 and 3 out of 10, for $N=1$ million, 10 million, 100 million, respectively, and the corresponding max constraint violation ratio is 2.5\%, 4.1\% and 3.0\%, respectively. Comparatively, the distributed SCD solutions have no violations. It is also worthwhile noting that the primal objective of pre-solving solution, even with constraint violations, is always smaller than the SCD solution.

\subsection{Scalability}
\begin{figure}
\centering
\includegraphics[width=0.725\linewidth]{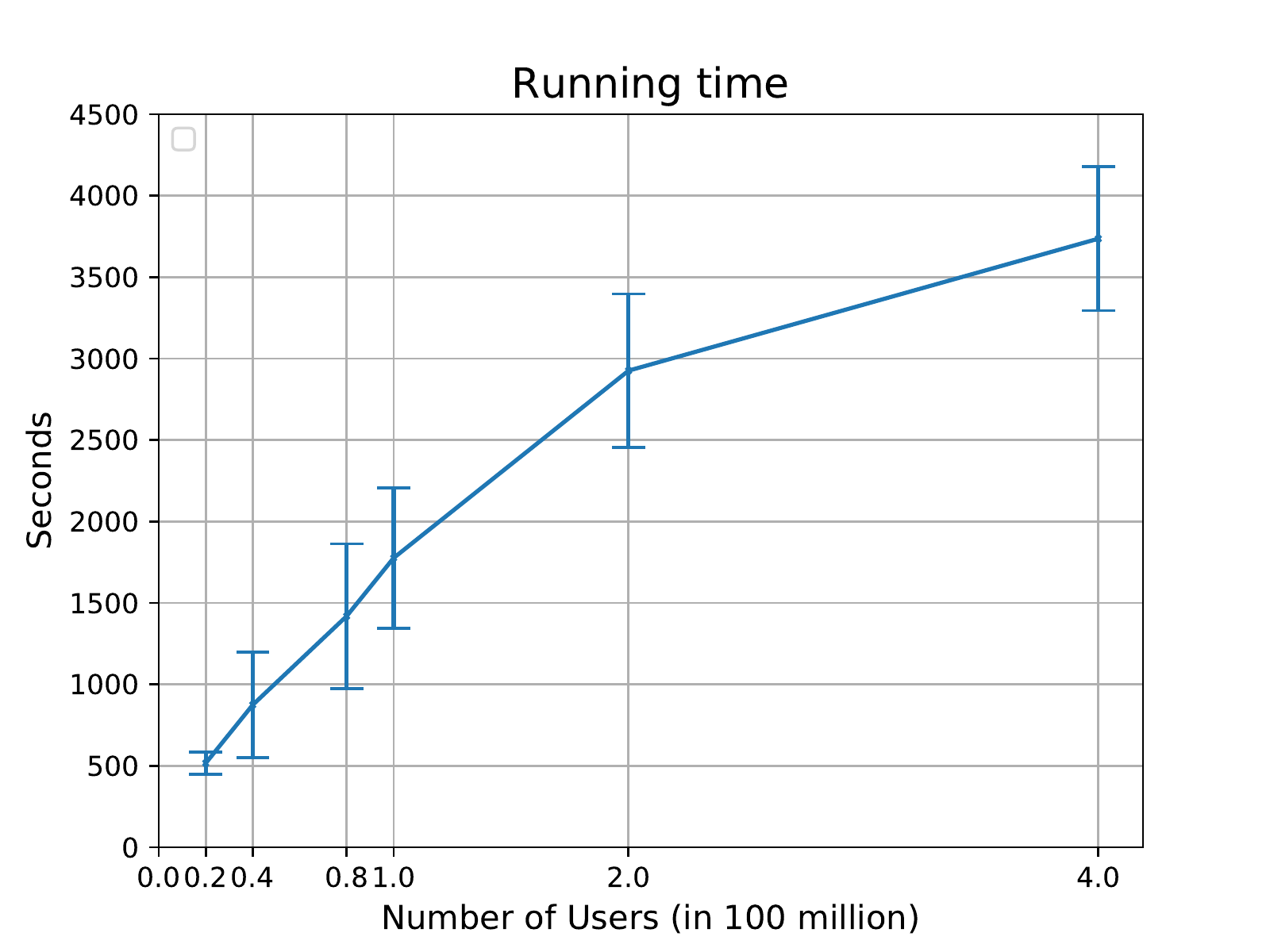}
\caption{Running time with $N=20, 40, 80, 100, 200, 400$ million users on 10 dense global constraints and 200 executors}
\label{ScalabilityN}
\end{figure}

\begin{figure}
\centering
\includegraphics[width=0.725\linewidth]{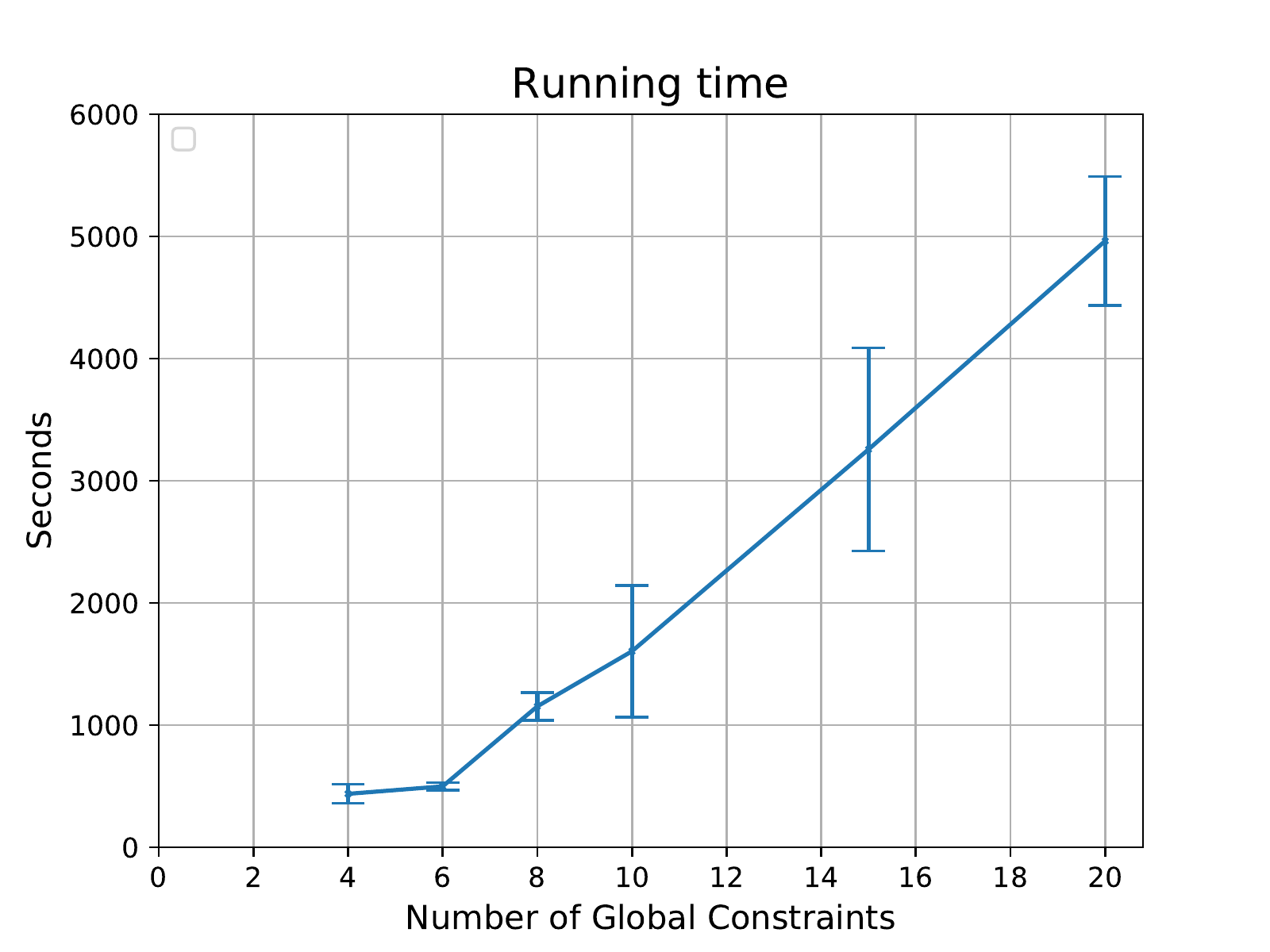}
\caption{Running time with $K=4, 6, 8, 10, 15, 20$ global dense constraints on 100 million users}
\label{ScalabilityK}
\end{figure}

\begin{figure}
\centering
\includegraphics[width=0.75\linewidth]{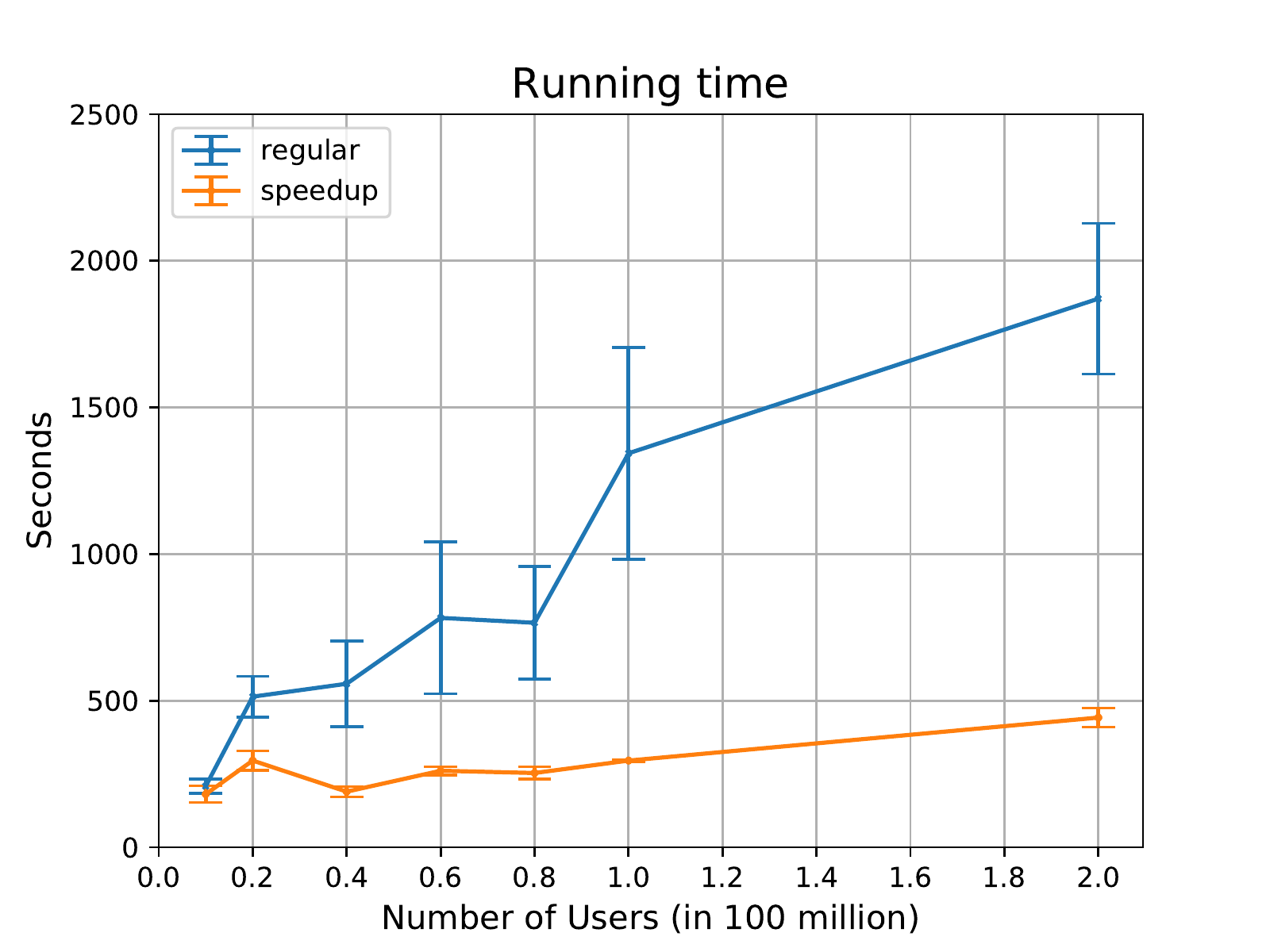}
\caption{Running time of speedup algorithm (speedup) and generalized algorithm (regular)}.
\label{ScalabilitySpeedup}
\end{figure}

To study the scalability of our solver, an implementation of our algorithm in Spark was tested using sparse and dense problem instances. Each Spark executor has 8 cores and 16G memory, and the number of executor is 200. Figure \ref{ScalabilityN} shows running time with $N=20, 40, 80, 100, 200, 400$ million users with $K=10$ dense global constraints and hierarchical local constraints. Figure \ref{ScalabilityK} plots running time with $K=4, 6, 8, 10, 15, 20$ global constraints, with $N$ fixed at $100$ million. Figure \ref{ScalabilityN} and \ref{ScalabilityK} illustrate that Algorithm \ref{AlgorithmSCD} is clearly scalable. Figure \ref{ScalabilitySpeedup} demonstrates that speedup algorithm in Section \ref{SectionSpeedup} reduces the running time significantly and consistently across different number of users with 10 global constraints.

The running time of our system has satisfied the business need of daily optimization of the decision variables. For example, when running on Spark with 200 executors in a shared Apache Hadoop computing infrastructure, the optimization for 1 billion decision variables and constraints was able to converge within an hour.


\subsection{Comparison of DD and SCD} \label{dd_scd}
\begin{figure}
\includegraphics[width=0.88\linewidth]{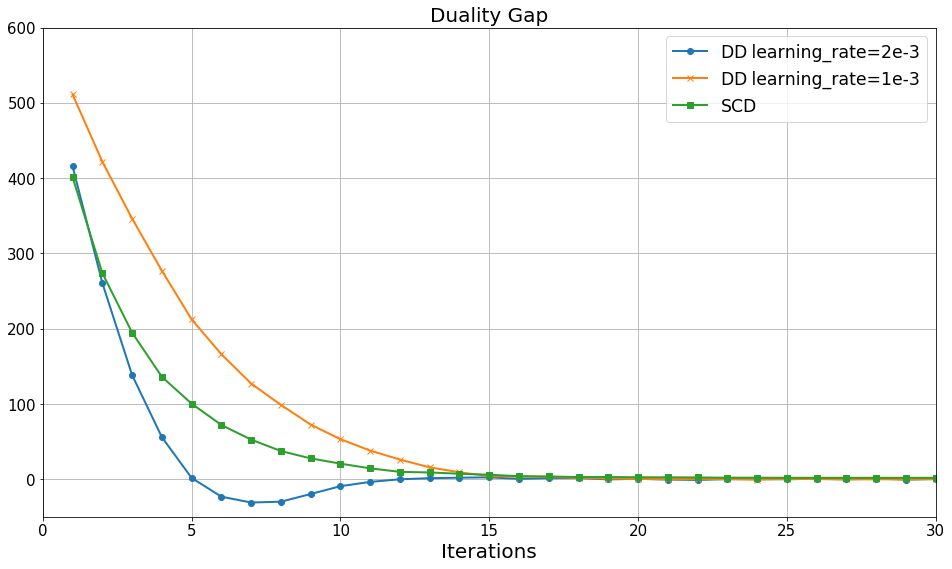}
\caption{Duality gaps for DD and SCD}
\label{FigureDualityGap}
\end{figure}

\begin{figure}
\includegraphics[width=0.88\linewidth]{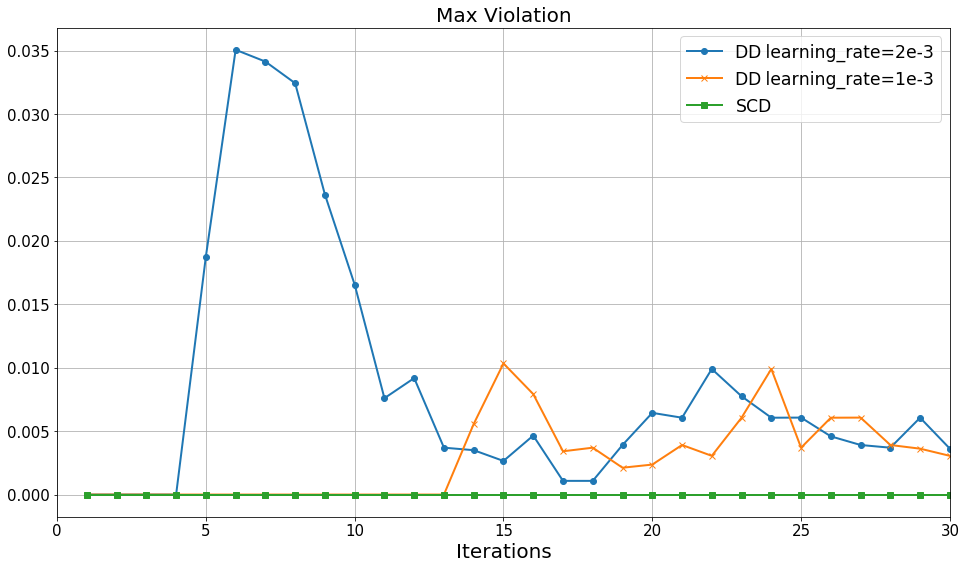}
\caption{Max constraint violation ratios for DD and SCD}
\label{FigureMaxViolation}
\end{figure}

To compare DD and SCD, both algorithms were run on sparse problem instances with $N=10,000$, $M=10$ and $K=10$. We experimented with a range of learning rates for DD, and present the results of learning rates being 1e-3 and 2e-3 here, as their convergence was most comparable to SCD in our experiment setting. Figure \ref{FigureDualityGap} plots the duality gaps as the number of iterations increases for the algorithms, while Figure \ref{FigureMaxViolation} shows the corresponding max constraint violation ratios. The results show that the number of iterations taken by both algorithms are comparable, but for SCD the max constraint violation ratio is much smaller and also way more smooth. SCD also requires less parameter tuning effort across problem instances, and is thus used for our real-world jobs.

\subsection{Production Deployment}
Ant Financial provides its users a wide spectrum of financial products and services ranging from payment to banking, loans, wealth management, insurances and so on. As of late 2018, Ant is serving over 1 billion active users globally through its mobile payment Apps such as Alipay. Solving KPs at scale is crucial to our business as various of financial resources are being allocated among our users on a daily basis. The distributed algorithms we developed in this paper were deployed to production in late 2018 and ever since have bee used to power production decision making everyday for more than 10 of our core products, e.g., marketing budget allocation, insurance pricing, online traffic control, notification volume optimization, credit risk management, and loan allocation, etc.

\section{Conclusions}
We introduce distributed algorithms for solving billion-scale knapsack problems. Our approach is developed based on a slightly generalized formulation of KPs and hence can be applied to solve other variants of KPs. The proposed algorithms can be easily implemented using common distributed frameworks such as MPI, Hadoop and Spark. The approach can also be extended to solve KPs with non-linear objective functions as long as it is decomposable with respect to the decision variables (or groups of decision variables).



\bibliographystyle{plain}
\bibliography{sample-bibliography}

\end{document}